\documentclass[aps,pra,twocolumn,superscriptaddress,nofootinbib]{revtex4}%
\usepackage{amssymb}
\usepackage{graphicx}
\usepackage{amsmath}
\usepackage{amsbsy}
\usepackage{cancel}
\usepackage{graphicx} 
\usepackage{dcolumn} 
\usepackage{bm} 
\usepackage[dvipsnames]{xcolor}
\usepackage{bbm}
\usepackage{exscale}
\usepackage{longtable}
\usepackage{braket}
\usepackage{empheq} 
\usepackage{enumerate}
\usepackage[mathscr]{euscript}

\usepackage{mathtools}

\usepackage{amsfonts}
\usepackage{amssymb,dsfont,physics}
\usepackage{tikz}
\usepackage{appendix}
\usetikzlibrary{decorations,snakes}
\usepackage{accents}%
\usepackage{multirow}
\usepackage{booktabs}
\providecommand{\U}[1]{\protect\rule{.1in}{.1in}}

\AtBeginDocument{
\heavyrulewidth=.08em
\lightrulewidth=.05em
\cmidrulewidth=.03em
\belowrulesep=.65ex
\belowbottomsep=0pt
\aboverulesep=.4ex
\abovetopsep=0pt
\cmidrulesep=\doublerulesep
\cmidrulekern=.5em
\defaultaddspace=.5em
}
\begin{document}
\title{{Optimal quantum metrology of two-photon absorption}}
\author{Athena Karsa}
\email{athena.karsa@gmail.com}
\affiliation{School of Physics \& Astronomy, University College London, London WC1E 6BT, UK}
\affiliation{Korea Research Institute of Standards and Science, Daejeon 34113, Republic of Korea}

\author{Ranjith Nair}
\affiliation{School of Physical and Mathematical Sciences, Nanyang Technological University, Singapore}

\author{Andy Chia}
\affiliation{Centre for Quantum Technologies, National University of Singapore, Singapore}

\author{Kwang-Geol Lee}
\affiliation{Department of Physics, Hanyang University, Seoul 04763, Republic of Korea}

\author{Changhyoup Lee}
\email{changhyoup.lee@gmail.com}
\affiliation{Korea Research Institute of Standards and Science, Daejeon 34113, Republic of Korea}

\date{\today}

\begin{abstract}

Two-photon absorption (TPA) is a nonlinear optical process with wide-ranging applications from spectroscopy to super-resolution imaging. Despite this, the precise measurement and characterisation of TPA parameters are challenging due to their inherently weak nature. 
We study the potential of single-mode quantum light to enhance TPA parameter estimation through the quantum Fisher information (QFI). Discrete variable (DV) quantum states (defined to be a finite superposition of Fock states) are optimised to maximise the QFI for given absorption, revealing a quantum advantage compared to both the coherent state (classical) benchmark and the single-mode squeezed vacuum state. For fixed average energy $\bar{n} \in 2\mathds{N}$, the Fock state is shown to be optimal for large TPA parameters, while a superposition of vacuum and a particular Fock state is optimal for small absorption for all $\bar{n}$. This differs from single-photon absorption where the Fock state is always optimal. Notably, photon counting is demonstrated to offer optimal or nearly optimal performance compared to the QFI bound for all levels of TPA parameters for the optimised quantum probes. 
Our findings provide insight into known limiting behaviours of Gaussian probes and their different Fisher information (FI) scalings under photon counting ($\propto \bar{n}^2$ for squeezed vacuum states versus $\bar{n}^3$ for coherent states). The squeezed state outperforms coherent states for small TPA parameters but underperforms in the intermediate regime, becoming comparable in the large absorption limit. This can be explained through fundamental differences between behaviours of even and odd number Fock states: the former's QFI diverges in both large and small absorption limits, while the latter diverges only in the small absorption limit, dominating at intermediate scales. 

\end{abstract}

\maketitle

\section{Introduction}

The study and use of nonclassical states of light is ubiquitous in quantum technologies, providing a potential resource for achieving a quantum enhancement in tasks such as metrology~\cite{giovannetti2011advances,polino2020photonic}, imaging~\cite{genovese2016real} and spectroscopy~\cite{mukamel2020roadmap}. Of particular interest is their possible role within nonlinear light-matter interactions such as two-photon absorption (TPA), the subject of this work. Notably, the TPA probability for a squeezed light source scales linearly with the intensity of the light field, as opposed to the quadratic scaling observed with laser light~\cite{klyshko1982transverse,gea1989two,javanainen1990linear,georgiades1995nonclassical,georgiades1997atoms,dayan2004two}. This effect may potentially enable nonlinear spectroscopy and microscopy at low photon fluxes which could in turn permit TPA-based protocols to interrogate photosensitive samples, or even living organisms, while mitigating their degradation~\cite{dorfman2016,schlawin2017,schwalin2018,gilaberte2019,ma2021,eshun2022}. 


Recent years have seen a multitude of ongoing developments in quantum-enhanced optical sensing protocols, owing to theoretical and experimental progress providing access to new quantum sources and means for their use~\cite{pirandola2018advances,karsa2020generic}. Concerning quantum-enhanced absorption measurements, the regime of single-photon absorption was naturally explored first~\cite{monras2007optimal,adesso2009optimal}. The Fock state at any fixed photon number ($\bar{n}$) was shown to saturate the ultimate bound of precision (maximised over all single-mode probes with average photon number $\bar{n}$) given by $1/4\bar{n}$ for a single run~\cite{
braunstein1996generalized,maccone2006information}, independent of the amount of absorption.


More recently, interest has been targeted towards TPA with focus primarily on the quantum metrological properties of Gaussian states~\cite{munoz2021quantum,panahiyan2022two}. Their expression in terms of a unitary operation acting on the vacuum allows for a simplification of the TPA analysis that, however, is valid only in the regime of extremely low TPA rates. While it is possible to consider this regime a useful one, since typical TPA cross-sections are generally small, this approach fails to allow for the study of arbitrary quantum states and a global comparison of their metrological potential across all possible scales of absorption.

\begin{figure}
    \centering
    \includegraphics[width=0.75 \linewidth]{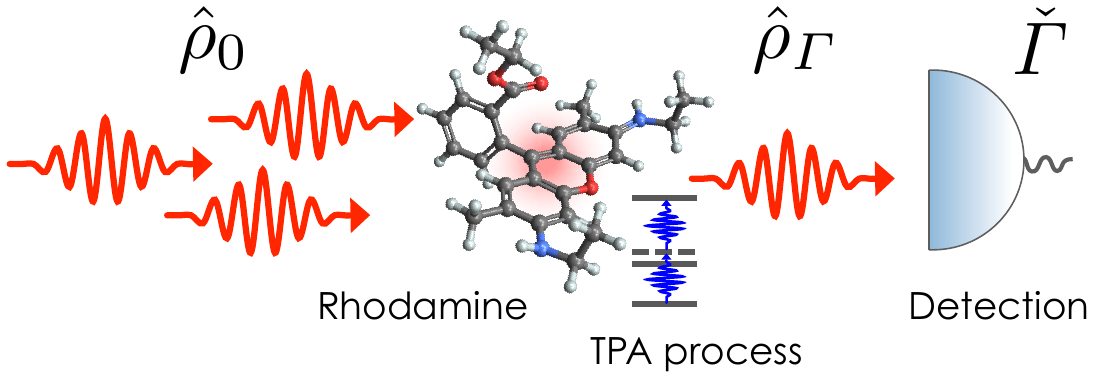}
    \caption{Schematic diagram of two-photon absorption (TPA) measurement. An input quantum state of light $\hat{\rho}_0$  illuminates a medium comprising TPA molecules (e.g., rhodamine) for a time $t$ which induce an incoherent TPA process with interaction strength conveniently captured by the dimensionless parameter $\varGamma = 1- \exp(-\gamma t)$, where $\gamma$ is the TPA rate, taking values in $[0,1)$ (c.f. Eq.~(\ref{eq:reparam})). After the interaction, the resultant output state $\hat{\rho}_\varGamma$ is measured by a detector, followed by post-processing to generate an estimate $\check{\varGamma}$.
    }
    \label{fig:TPAdiagram}
\end{figure}

This work studies the metrological power of various bosonic single-mode states $\hat{\rho}_0$ for estimating TPA rates as modelled by (illustrated in Fig.~\ref{fig:TPAdiagram}),  
\begin{align}
\label{OriginalMasterEqn1}
\frac{d}{dt}\,\hat{\rho}_t = \gamma \, \mathscr{L} \, \hat{\rho}_t  \;,
\end{align}
where $\gamma$ is the TPA rate to be estimated, and $\mathscr{L}$ is a superoperator in the Lindblad form, defined for any $\hat{J}$ and $\hat{s}$ by~\cite{wiseman2010quantum}
\begin{align}
\label{OriginalMasterEqn2}
       \mathscr{L} \, \hat{s} \equiv \hat{J} \, \hat{s} \, \hat{J}^{\dagger} - \frac{1}{2} \, \hat{J}^{\dagger} \hat{J} \, \hat{s} - \frac{1}{2} \, \hat{s} \, \hat{J}^{\dagger} \hat{J} \;.
\end{align}
For TPA, we take $\hat{J}=\hat{a}^2/\sqrt{2}$ with $\hat{a}$ being the bosonic annihilation operator. We have set $\hbar=1$ and moved into a rotating frame in which the the free harmonic evolution of the probe field appears stationary.

Our primary aim is to find, for a fixed average photon number $\bar{n}$, the probe state $\hat{\rho}_0$ which maximises the quantum Fisher information (QFI) associated with estimating $\gamma$. We find that Fock states with $\bar{n} \in 2\,\mathds{N}$ are optimal only in the limit of very large TPA, with competing behaviour between squeezed vacuum and coherent states at the low to intermediate TPA scales. This suggests an absorption-dependent behaviour in the form of the optimal probe. With this in mind, we optimise a general discrete variable (DV) quantum state of the form 
\begin{equation}
    \ket{\psi_{\mathrm{DV}}} = \sum_{j=0}^{N_{\mathrm{max}}} c_j \ket{j} \;,
    \label{eq:genDVstate}
\end{equation}
to maximise the QFI given a fixed TPA parameter value, where $\ket{j}$ is a Fock state with photon number $j$. The theoretically optimal single-mode quantum probe is found and our analysis also provides insight into the observed behaviours of the coherent, squeezed vacuum and Fock states, to which our probe is compared. While the Fock state is not always optimal, superpositions of the vacuum with another Fock state are, where the photon number of the latter state is a function of the TPA parameter.

\section{Background}\label{sec:QuantumMetrologyTPA}

Aspects of parameter estimation theory are briefly reviewed here. We also discuss the TPA model, including its assumptions and the computations that lead to the results of Sec.~\ref{Results}.



\subsection{Quantum parameter estimation theory}

The goal of parameter estimation is to determine the value of some unknown parameter $\varepsilon$ (which we take to be deterministic) given a set of observed data, say $\bm{y} \equiv (y_1, y_2, \ldots, y_\nu)$ (which we assume to be discrete).  The data $\bm{y}$ are outcomes of measurements drawn from the conditional probability $p(\bm{y}|\varepsilon)$, conditioned on the true value of $\varepsilon$. An estimator $\check{\varepsilon}(\bm{y})$ is then constructed from $\bm{y}$ to form an inference of $\varepsilon$. For an unbiased estimator, i.e.,~one for which $\langle\check{\varepsilon}\rangle=\varepsilon$, classical estimation theory bounds the mean-squared error of $\check{\varepsilon}$ by the Cram\'{e}r--Rao (CR) inequality 
\begin{equation}
    \langle ( \check{\varepsilon} - \varepsilon )^2 \rangle \geq \frac{1}{\nu \mathcal{F}} \; ,
    \label{CRB}
\end{equation}
where $\langle g(\bm{y}) \rangle = \sum_{\bm{y}} \, p(\bm{y}|\varepsilon) \, g(\bm{y})$ for any $g(\bm{y})$. The bound is determined by the Fisher information (FI), defined as
\begin{align}
\label{eq:CFI}
        \mathcal{F} = \sum_{\bm{y}} \, p(\bm{y}|\varepsilon) \, \bigg[ \frac{\partial}{\partial\varepsilon} \, \ln p(\bm{y}\vert\varepsilon)\bigg]^2 \;.
\end{align}
Quantum parameter estimation theory is necessary when the measurement yielding $\bm{y}$ becomes a quantum process, which is described by a positive operator-valued measure (POVM),
\begin{align}
    \mathbbm{P} = \bigg\{ \hat{\Pi}_{\bm{y}} \,\bigg| \,\hat{\Pi}_{\bm{y}}\ge0, \, \sum_{\bm{y}}\hat{\Pi}_{\bm{y}}=\hat{\mathds{1}} \bigg\}  \; .
\end{align} 
The FI retains its form as in Eq.~\eqref{eq:CFI}, but the probability of observing $\bm{y}$ is now given by $p(\bm{y}|\varepsilon)=\Tr\!\big[\hat{\rho}_\varepsilon\hat{\Pi}_{\bm{y}}\big]$, where $\hat{\rho}_\varepsilon$ is the state of the system carrying $\varepsilon$. A further reduction of the CR bound is then possible, expressed by the well-known quantum CR (QCR) inequality \cite{helstrom1967minimum},
\begin{equation}
    \frac{1}{\mathcal{F}} \geq \frac{1}{\mathcal{F}_\text{Q}} \,.
    \label{QCRB}
\end{equation}
The quantity $\mathcal{F}_\text{Q}$ is the quantum Fisher information (QFI), defined by
\begin{align}
\label{DefnQFI}
    \mathcal{F}_\text{Q} = \Tr\big[ \hat{\rho}_\varepsilon \hat{L}^2_\varepsilon \big]  \;,
\end{align}
where $\hat{L}_{\varepsilon}$ is the symmetric logarithmic derivative (SLD), defined as a Hermitian solution to 
\begin{equation}
    \frac{d \hat{\rho}_{\varepsilon}}{d {\varepsilon}} = \frac{1}{2}\,\big( \hat{L}_{\varepsilon} \hat{\rho}_{\varepsilon} + \hat{\rho}_{\varepsilon} \hat{L}_{\varepsilon} \big)\;.
    \label{DefnSLD}
\end{equation}
It is straightforward to show from Eq.~\eqref{DefnSLD} that
\begin{equation}
    \hat{L}_{\varepsilon} = 2 \underset{(\lambda_k+\lambda_l > 0)}{\sum_{k} \sum_{l}} \frac{\bra{\lambda_l} (d\hat{\rho}_{\varepsilon}/d\varepsilon) \ket{\lambda_k}}{\lambda_{k} + \lambda_{l}} \dyad{\lambda_l}{\lambda_k},
        \label{eq:SLD}
\end{equation}
where $\hat{\rho}_\varepsilon \ket{\lambda_k} = \lambda_k \ket{\lambda_k}$, and note that $\lambda_k$ and $\ket{\lambda_k}$ are $\varepsilon$-dependent. The QFI in Eq.~\eqref{DefnQFI} arises as a maximisation of the FI in Eq.~\eqref{eq:CFI} over all possible measurements. That is, $\mathcal{F}_\text{Q}=\max_\mathbbm{P} \mathcal{F}$, and the optimal $\mathbbm{P}$ can be constructed over the eigenbasis of $\hat{L}_\varepsilon$ \cite{braunstein1994statistical}.

\subsection{TPA model and computations}

It is convenient to nondimensionalise Eq.~\eqref{OriginalMasterEqn1} by using the dimensionless parameter $\varepsilon=\gamma\,t$. Our model for TPA then becomes
\begin{equation}
    \label{TPAmastereqn}
        \frac{d}{d\varepsilon} \, \hat{\rho}_\varepsilon = \mathscr{L} \, \hat{\rho}_\varepsilon \;.
\end{equation} 
This reparametrisation also makes sense physically because the total number of TPA events in a given sample can be achieved by controlling how much time the probe light interacts with the sample. Note that we have assumed the molecules or atoms of the TPA sample to be in their ground state. This ensures the only possible two-photon transitions in the sample are absorption processes, i.e.~no two-photon emissions occur. Two-photon emission from sample molecules may be accounted for by another Lindblad superoperator with $\hat{J}=\hat{a}^\dag{}^2/\sqrt{2}$, which lead to other nontrivial effects \cite{lambropoulos1967quantum,chia2020phase,chia2023quantum}. As in Ref.~\cite{munoz2021quantum}, one-photon transitions in the TPA sample are neglected.

The exact analytical solution $\hat{\rho}_\varepsilon$ of Eq.~\eqref{TPAmastereqn} for an arbitrary input state $\hat{\rho}_0$ has been obtained in the literature~\cite{simaan1975quantum, simaan1975quantum2, simaan1978off, gilles1993two, klimov2003algebraic}. However, the exact solution from the literature is rather complicated and we resort to numerical computations of the QFI. Even so, the arbitrariness of $\hat{\rho}_0$ permitted by the exact solution is still important, as it affords us a way to find states $\hat{\rho}_0$ that maximise the QFI. The exact solution is explained in Appendix~\ref{App1}. To calculate the QFI we need to calculate first the SLD as defined in Eq.~\eqref{eq:SLD}, which in turns requires $\lambda_k$ and $\ket{\lambda_k}$. For each value of $\varepsilon$, the output state $\hat{\rho}_{\varepsilon}$ is numerically evaluated using Eq.~(\ref{KlimovSolnFinal}) from Appendix~\ref{App1}, and then diagonalised to find $\lambda_k$ and $\ket{\lambda_k}$. This then allows us to evaluate $\bra{\lambda_l}(d\hat{\rho}_{\varepsilon}/d\varepsilon) \ket{\lambda_k}$ using Eqs.~\eqref{OriginalMasterEqn2} and \eqref{TPAmastereqn} with $\hat{J}=\hat{a}^2/\sqrt{2}$. The SLD is then constructed according to Eqs.~\eqref{OriginalMasterEqn2} and \eqref{eq:SLD}, from which the QFI can be readily calculated using Eq.~\eqref{DefnQFI}.

While we carry out our calculations with the variable $\varepsilon$, it is more convenient to consider the final results (such as the QFI) in terms of
\begin{align}
\label{eq:reparam}
    \varGamma = 1 - e^{-\varepsilon}  \; .
\end{align}
This is because $\varepsilon$ is unbounded, taking on values in $[0,\infty)$, whereas $\varGamma$ has values restricted to $[0,1)$. The finite range of $\varGamma$ is helpful for plotting and visualisation. Note that $\varepsilon$ is in one-to-one correspondence with $\varGamma$. If $\varepsilon$ increases (or decreases), $\varGamma$ increases (or decreases). Thus no ambiguity arises in using $\varGamma$ to discuss our results.\footnote{We do not recast Eq.~\eqref{OriginalMasterEqn1} in terms of $\varGamma=1-\exp(-\gamma t)$ at the outset because this results in an explicitly $\varGamma$-dependent equation. The solution to such an equation then becomes very difficult to evaluate.}

\section{Results and Discussion}
\label{Results}






\subsection{TPA parameter estimation under an energy constraint}


\begin{figure}
    \centering
    \includegraphics[width=\linewidth]{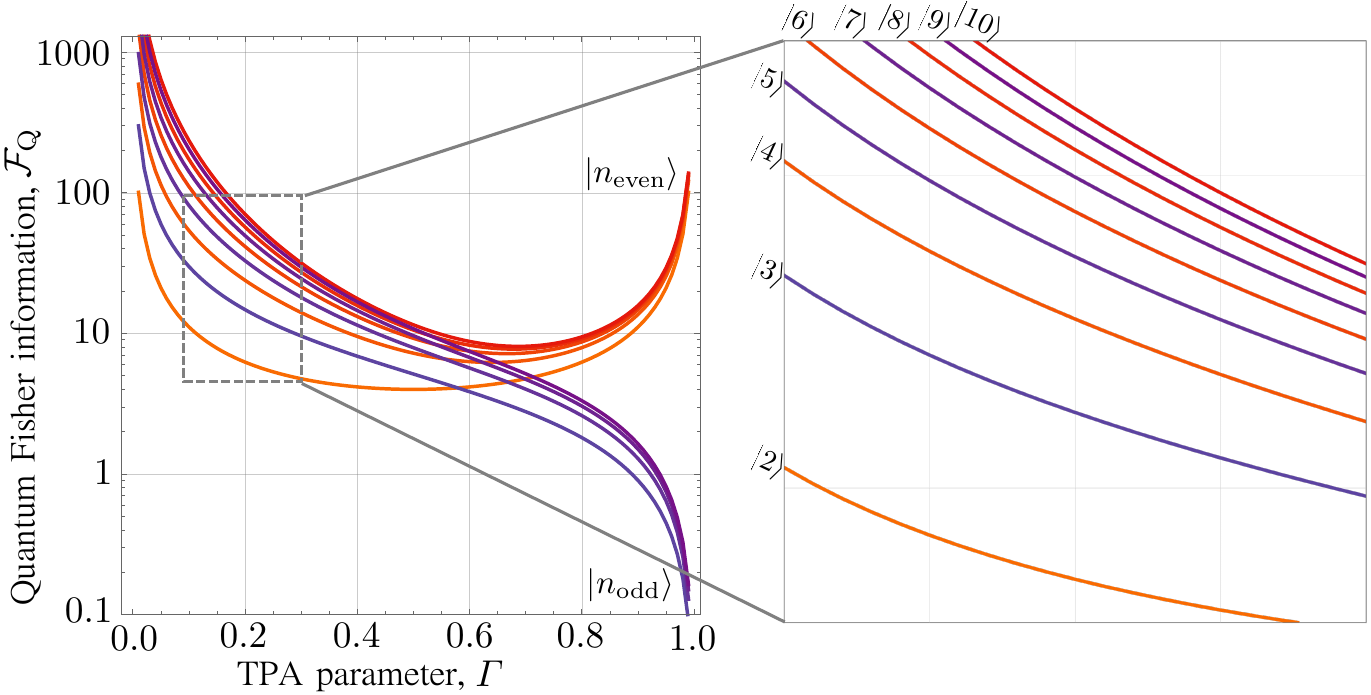}
    \caption{QFI for TPA parameter estimation for even and odd Fock states where $n_{\rm even}=0,2,4,\ldots,$ and $n_{\rm odd}=1,3,5,\ldots.$ Clearly shown are differing behaviours across $\varGamma$, which will explain the behaviours of the other probe states to be considered in this work.}
    \label{fig:evenvsodd}
\end{figure}

\subsubsection{Fock states}

In the case of single-photon absorption estimation, with Lindblad operator $\hat{J} = \hat{a}$, Fock states at any fixed photon number are known to unconditionally saturate the ultimate bound on precision~\cite{adesso2009optimal, braunstein1996generalized}. In the case of TPA parameter estimation, on the other hand, this does not hold for all values of mean photon numbers and TPA parameters. In particular, there exists a dramatic difference in the behaviours of even and odd Fock states subject to TPA evolution, and their resultant precision to TPA parameter estimation. This contrast is illustrated in Fig.~\ref{fig:evenvsodd}. For weak TPA, Fock states $\ket{n}$ with higher $n$ have higher QFIs, all diverging as $\varGamma \rightarrow 0$. In this limit, the scaling of the QFI for a Fock state with photon number $\bar{n}$ goes as (see Appendix~\ref{App3} for the derivation),
\begin{equation}
    \mathcal{F}_{\mathrm{Q},\mathrm{F}} \simeq \frac{\bar{n}(\bar{n} -1)}{2 \varGamma} \;.
    \label{eq:fockscalingsmallTPA}
\end{equation}
In the limit of strong TPA, QFIs for odd Fock states decay to zero while those for even Fock states diverge once again. The performances of Fock states with $\bar{n}=2$ and $\bar{n}=3$ will be compared with other single-mode probe states of the same mean energy in the next section.

\subsubsection{Optimal DV, squeezed vacuum and coherent states}\label{sec:optimalTPA}

Since the QFI depends only on the input state $\hat{\rho}_{0}$, it can be maximised when the input state is chosen appropriately. In this subsection, we maximise the QFI over a large class of input states subject to the global constraint of fixed mean photon number $\bar{n}$. 

An optimisation is performed over all possible DV quantum states which take the form of Eq.~(\ref{eq:genDVstate}) for a given maximum occupation number $N_{\mathrm{max}}$, and a fixed mean photon number $\bar{n}$. For simplicity, the coefficients $c_j$ are assumed to be real and positive, and we set $N_{\mathrm{max}}=10$. A nested optimisation procedure is employed, comprising two successive algorithms: the evolutionary algorithm library EvoTorch~\cite{evotorch2023arxiv} is used for a coarse-grained global search of maxima, and then its output forms the seed of a subsequent adaptive moment estimation optimiser (called Adam) which fine-tunes the result through gradient descent. This procedure is carried out for each value of $\varGamma$, yielding paired results comprising the maximal QFI and the optimal DV quantum state attaining that QFI.

The QFI corresponding to different optimal DV states are compared with those for squeezed vacuum states $\vert \xi\rangle=\hat{S}(\xi)\vert 0\rangle$ with a squeezing operator $\hat{S}(\xi)=\exp\!\big[(\xi^*\hat{a}^2 -\xi \hat{a}^{\dagger2})/2\big]$, where~$\xi=r \exp(i\theta_{\rm s})$ and~$r\ge0$; Fock states $\vert n \rangle$; and coherent states $\vert \alpha\rangle=\hat{D}(\alpha)\vert 0 \rangle$ with a displacement operator $\hat{D}(\alpha)=\exp(\alpha \hat{a}^{\dagger}-\alpha^* \hat{a})$ for $\alpha=\vert \alpha \vert \exp(i\theta_\text{d})$. The coherent state serves as a classical benchmark allowing us to assess the quantum advantage in using $\ket{\xi}$, $\ket{n}$, or $\ket{\psi_{\mathrm{DV}}}$ as probe states. For fair comparisons, a fixed mean photon number $\bar{n}$ is enforced by setting $\bar{n} = n = \sinh^2(r) = |\alpha|^2$. We also set $\theta_\text{s}=\theta_\text{d}=0$ for simplicity. This is permitted since the QFI is actually independent of these phases (see Appendix~\ref{App4}).

\begin{figure}
    \centering
    \includegraphics[width=\linewidth]{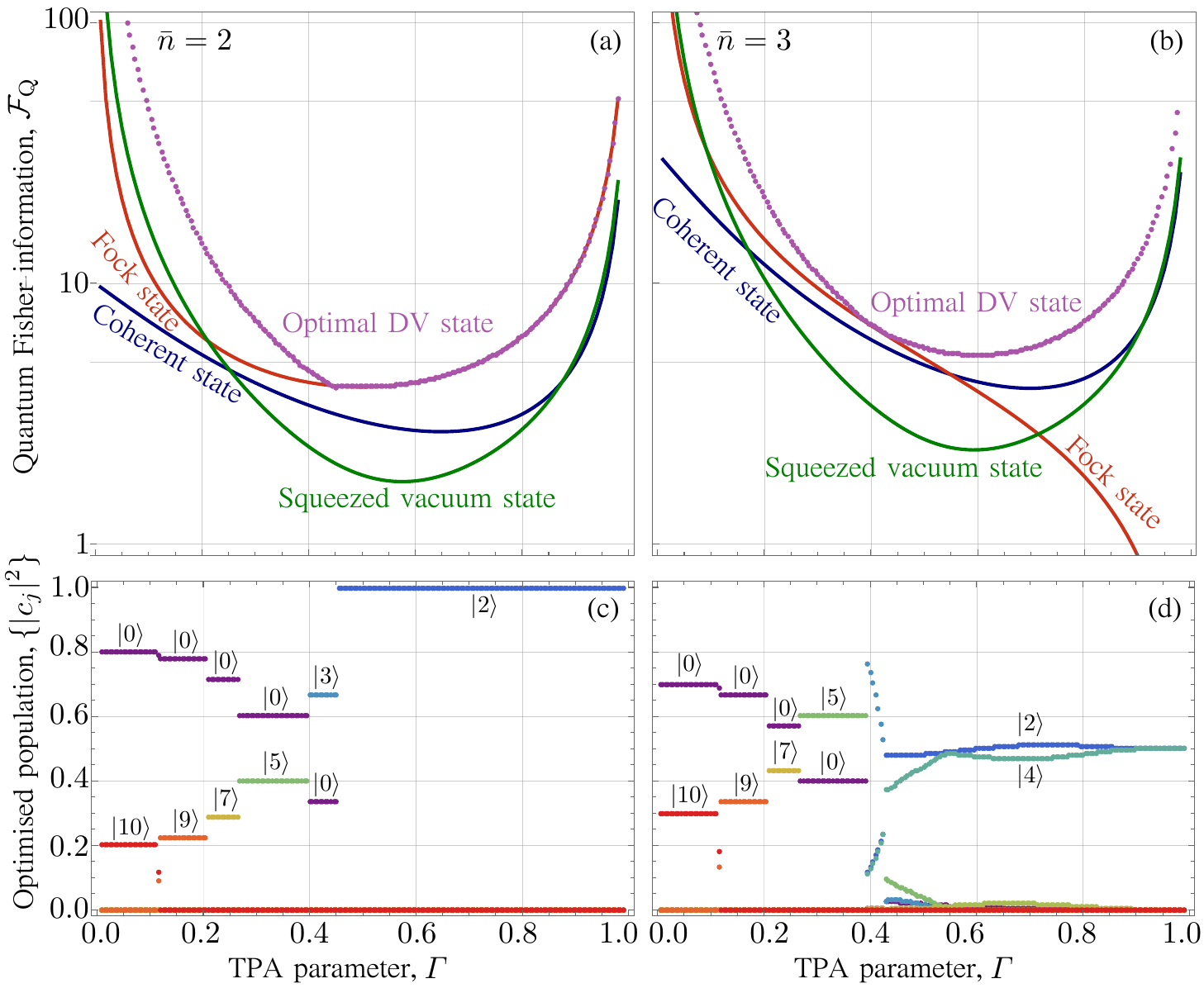}
    \caption{(a) and (b) QFI as a function of $\varGamma$ of the optimal DV probe state with $N_{\mathrm{max}}=10$, compared with that of the coherent, squeezed vacuum and pure Fock state for a mean number of photons $\bar{n} = 2$ and $3$, respectively. (c) and (d) The corresponding optimised DV state populations $\{|c_j|^2\}$, yielding optimised DV probe $\ket{\psi_{\mathrm{DV}}} = \sum_{j=0}^{N_{\mathrm{max}}} c_j \ket{j}$ attaining the maximum QFI for TPA parameter estimation, across all $\varGamma$. Note that for each value of $\varGamma$, all values of $\{|c_j|^2\}$ with $j=0,\dots 10$ are plotted and coloured according to the rainbow across the range. For most data sets, all coefficients are zero except those corresponding to the vacuum and the optimal $\ket{N}$ for the $\varGamma$ in question.}
    \label{fig:QFIoptstaten2and3}
\end{figure}

The QFIs resulting from the optimisation are depicted as a function of $\varGamma$ in Figs.~\ref{fig:QFIoptstaten2and3}(a) and (b) for mean photon numbers $\bar{n}=2$ and 3, respectively. The QFIs for optimal DV states are dominant across the entire range of $\varGamma$ while the squeezed vacuum, Fock, and coherent states exhibit competitive behavior between themselves. For $\bar{n}=2$, the squeezed vacuum state surpasses the other two states for small $\varGamma$, but the Fock state, i.e., $\vert 2\rangle$, starts to be dominant when $\varGamma \gtrsim 0.2$ and coincides with the optimal DV states when $\varGamma \gtrsim 0.45$. While the Fock state provides a quantum advantage over the coherent state across all $\varGamma$ values, the squeezed vacuum state's quantum advantage only exists at small or large $\varGamma$. The QFIs of all considered states diverge as $\varGamma\rightarrow0$ or $\varGamma\rightarrow1$, except for the coherent state whose QFI approaches a finite value of $\bar{n}^3 + \bar{n}^2/2$ as $\varGamma\rightarrow0$ (note that this value holds for all $\bar{n} \in \mathds{R}$). For $\bar{n}=3$, the overall behaviours are similar to the case of $\bar{n}=2$, except now the Fock state, i.e., $\ket{3}$, has a monotonically decreasing QFI as $\varGamma$ increases (which was also seen in Fig.~\ref{fig:evenvsodd}).


Figures~\ref{fig:QFIoptstaten2and3}(c) and (d) illustrate the populations $\{|c_j|^2\}_{j=0}^{N_{\mathrm{max}}}$ of the optimal DV states, with maximum occupation number $N_{\mathrm{max}}=10$, which resulted in the maximised QFIs of Figs.~\ref{fig:QFIoptstaten2and3}(a) and (b). For $\bar{n}=2$, we can see that the resultant optimal DV state can be written in the simple form of the weighted ON state $\vert \text{ON} \rangle=c_0\vert 0\rangle+c_N\vert N(\varGamma)\rangle$, where $N(\varGamma)$ denotes the optimal occupation found through optimisation for a given $\varGamma$, and the weights $c_{0}$, $c_N$ take values to yield the mean $\bar{n}$: $c_0 = \sqrt{1-c_N^2}$ and $c_N = \sqrt{\bar{n}/N(\varGamma)}$. 

Following an equivalent procedure as for the pure Fock state but taking into account the weighting $c_N$ when $\varGamma$ is small, it can be determined that as $\varGamma \rightarrow 0$, the QFI of this optimal probe is
\begin{equation}
    \mathcal{F}_{\mathrm{Q},\mathrm{ON}} \simeq \frac{\bar{n}(N-1)}{2 \varGamma}.
    \label{eq:ONscalingsmallTPA}
\end{equation}
In the case of $\bar{n}=3$, the optimal DV state can be written similarly in the form of the weighted ON state, but it evolves into a superposition of $\vert 2\rangle$ and $\vert 4\rangle$ when $\varGamma\gtrsim 0.45$. Across this region it can be found that there exists a wide range of states with similar QFIs, leading to multiple local optima and some transient points in Fig.~\ref{fig:QFIoptstaten2and3}(d). 

The differing behaviours across $\varGamma$ for even and odd Fock states manifests in the way the value of $N(\varGamma)$ of the optimal probe, in the form of ON state, varies across $\varGamma$ as well. In particular, it is apparent that there exists, for some value of $\varGamma$, a phase transition between odd and even number state dominance in terms of their associated QFIs. This is discussed further in Appendix~\ref{AppEvenOdd}.


The form of the optimal probe also provides further insight into the already known behaviours for coherent and squeezed vacuum states in the limit of $\varGamma \rightarrow 0$. In this regime, both the coherent state QFI and photon-counting FI saturates at $\bar{n}^3 + \bar{n}^2/2$ while the squeezed vacuum state's FI is $\sim 10\bar{n}^2$~\cite{munoz2021quantum}. For relatively low energies, $\bar{n} \lesssim 10$, the squeezed vacuum can obtain higher precision than the coherent state even with sub-optimal photon counting. At these energies, however, the heavy-tailed distribution of the squeezed vacuum state naturally emulates our proposed optimal probe, with a larger proportion of its population in higher-energy states, while the coherent state remains narrowly distributed around its mean, i.e., smaller populations in higher-energy states. As the mean energy is progressively increased, the optimal probe becomes closer in distribution to the pure Fock state which the coherent state's photon number distribution more closely resembles, allowing it to outperform the squeezed vacuum. 

This explanation can go further to also explain behaviours seen outside the limit of vanishing $\varGamma$: intermediate to high TPA parameters show the optimal probe comprising of Fock states becomes increasingly centered around its mean (with optimised $N \rightarrow \bar{n}$) with QFI contributions from odd number states becoming dominant; here the coherent state yields higher precision than the squeezed vacuum. Eventually, with the degradation of all odd number state contributions to zero and the divergence of even number state contributions, we see the squeezed vacuum's QFI recovering once again as $\varGamma \rightarrow 1$.

\begin{figure}
    \centering
    \includegraphics[width=\linewidth]{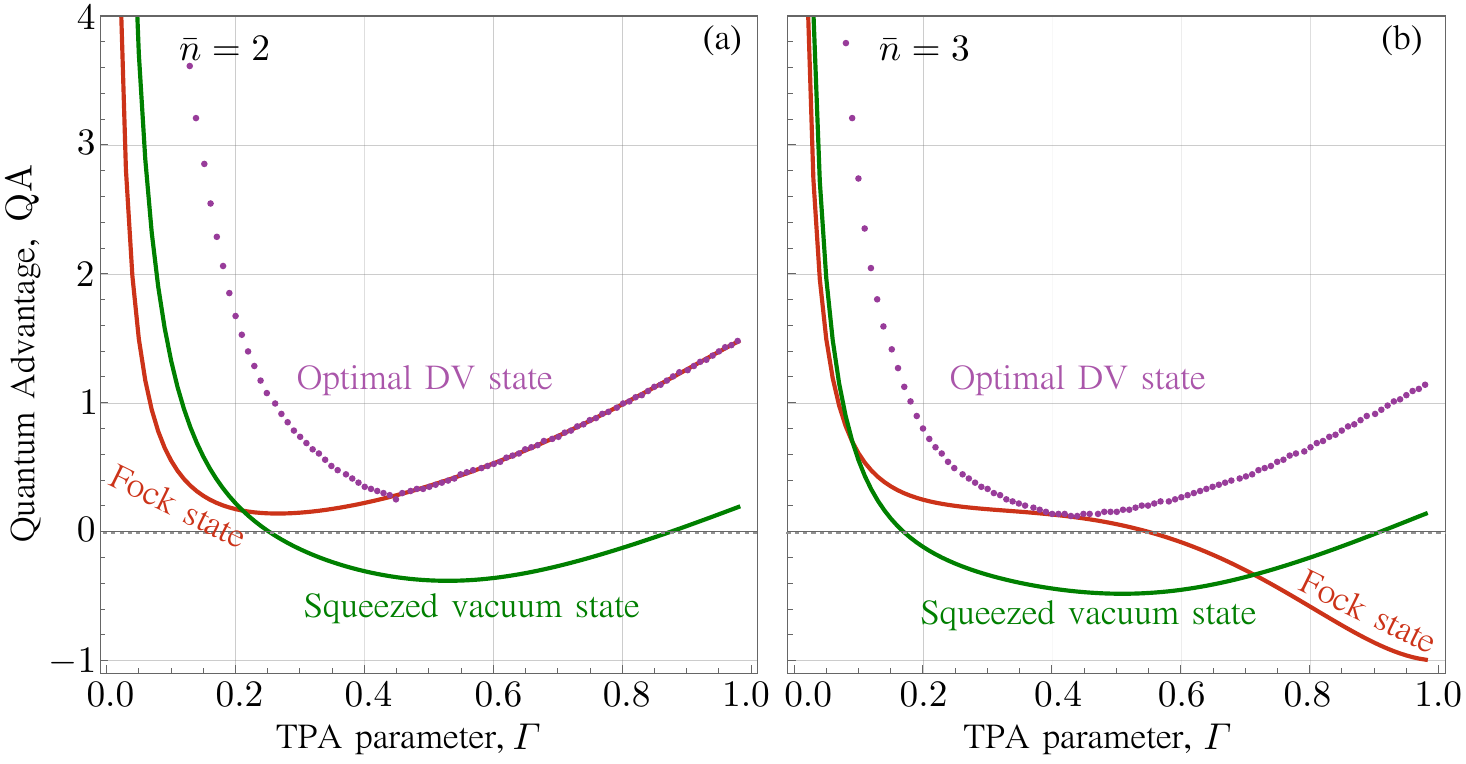}
    \caption{Quantum advantage, QA, of the squeezed vacuum, Fock and optimal DV state for TPA parameter estimation relative to the coherent state classical benchmark, quantified via their QFIs, for (a) $\bar{n}=2$, and (b) $\bar{n}=3$.}
    \label{fig:quantumadv}
\end{figure}

\subsection{Quantum advantage}

Figure~\ref{fig:quantumadv} shows the quantum advantage in TPA parameter estimation afforded by the use of the three quantum states relative to the coherent state benchmark: the squeezed vacuum, Fock, and optimal DV probe. The quantum advantage is defined as
\begin{equation}
    \mathrm{QA} = \frac{\mathcal{F}_{\mathrm{Q},\square} - \mathcal{F}_{\mathrm{Q}, \mathrm{Coh}}}{\mathcal{F}_{\mathrm{Q},\mathrm{Coh}}},
\end{equation}
where $\mathcal{F}_{\mathrm{Q},\square}$ is the QFI for the specific quantum state where we take $\square \in \{\mathrm{SV},\mathrm{F},\mathrm{opt}\}$, denoting the squeezed vacuum, Fock and optimal DV probe, respectively. 

For $\bar{n}=2$, the optimal DV probe and Fock state always show a quantum advantage while there is a region of intermediate $\varGamma$ where the squeezed vacuum probe loses its advantage. When $\bar{n}=3$, again the optimal DV probe consistently shows a quantum advantage while the Fock state eventually loses it in the limit of large $\varGamma$. The squeezed vacuum state, like before, has a quantum advantage for small $\varGamma$, losing it for intermediate absorption values before regaining it as $\varGamma\rightarrow 1$.

\subsection{Photon number detection}

\begin{figure}
    \centering
    \includegraphics[width=\linewidth]{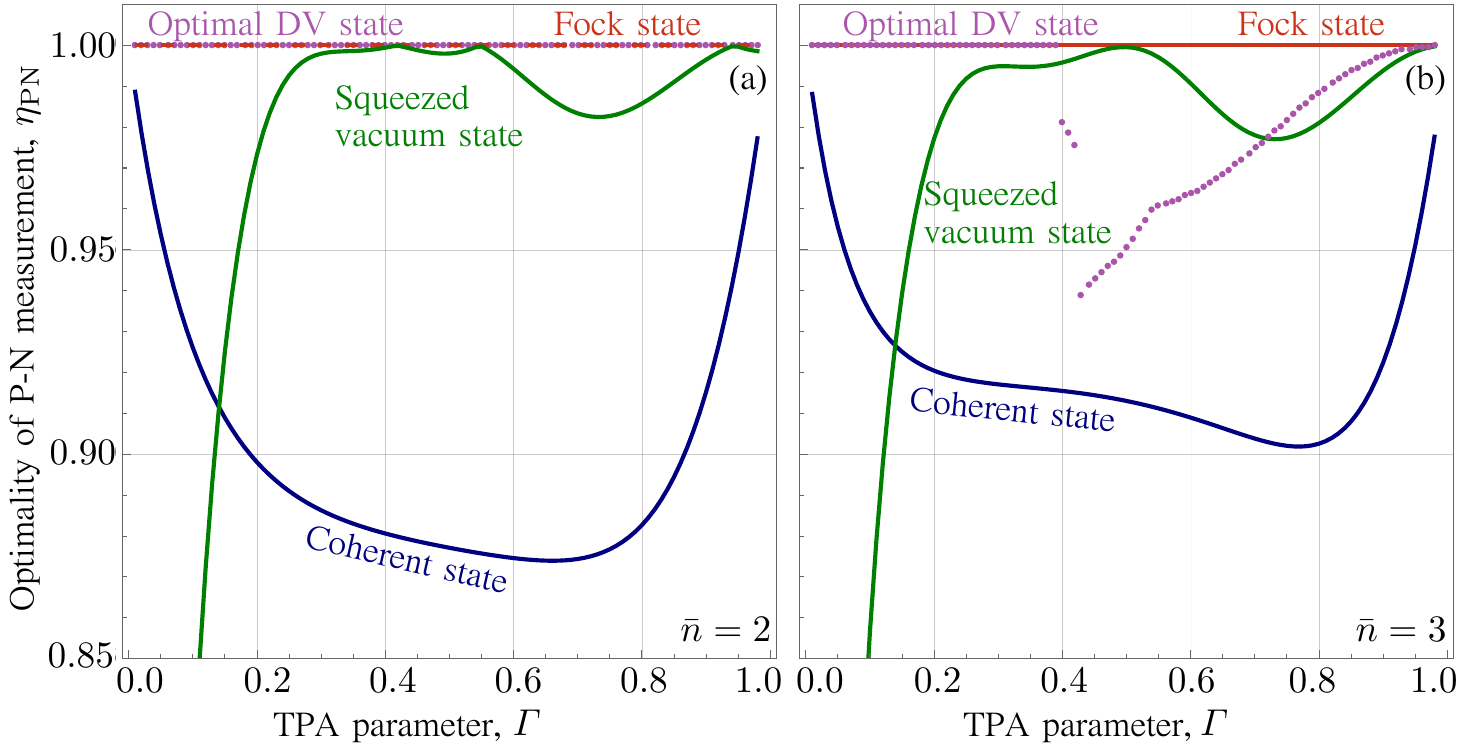}
    \caption{Efficiency of photon number (PN) measurements, $\eta_{\rm PN}$ for the squeezed vacuum, Fock, optimal DV and coherent state for TPA parameter estimation when (a) $\bar{n}=2$, and (b) $\bar{n}=3$.}
    \label{fig:photonnumberdelta}
\end{figure}

To show how good photon number measurements are for TPA parameter estimation using each of the quantum states considered in this work, we define the efficiency of photon counting as the following ratio
\begin{equation}
    \eta_{\rm PN} = \frac{ \mathcal{F}_{\rm PN}}{\mathcal{F}_{\mathrm{Q}}},
\end{equation}
where $\mathcal{F}_{\rm PN}$ is the FI due to photon-number (PN) measurements. The results are plotted in Fig.~\ref{fig:photonnumberdelta}.

For coherent states, photon counting is never completely optimal but improves in the limits of $\varGamma \rightarrow 0$ and $\varGamma \rightarrow 1$. In comparison, for a squeezed vacuum probe, photon counting is far from optimal when absorption is small and, overall, improves to becoming nearly optimal for increasing TPA. Meanwhile Fock state QFIs are, unsurprisingly, saturated by photon counting. 

Remarkably, when the optimal probe is of the form of the ON state, the optimal measurement that saturates the QFI is simple photon counting. There are, however, some subtle intricacies to this fact depending on the nature of $N$: Photon-counting is always the optimal measurement for ON states with $N = N_{\mathrm{o}} \in \mathds{O}$, where $\mathds{O}$ refers to the set of odd positive integers, while, for ON states with $N = N_{\mathrm{e}}\in \mathds{E}$, where $\mathds{E}$ is set of positive even integers photon-counting is not optimal but becomes increasingly close to optimal for increasing $N\geq 4 \in \mathds{E}$ and $\varGamma$.

The optimality of photon counting applies to all optimal probes for all values of $\varGamma$ whose mean energies $\bar{n} \in \mathds{E}$, where $\mathds{E}$ is the set of positive even integers. For such states, the optimal probes are all either odd ON states, for which the previous argument applies, or the pure, even Fock state after some value of $\varGamma$. When $\bar{n} \notin \mathds{E}$, there exists a region $\varGamma \gtrsim 0.4$ in which the transition occurs between odd and even Fock state dominance. Before this, the optimal probes are odd ON states and the optimality of photon counting still applies, but after this, the optimal probe becomes a superposition of the lowest-lying even number states and photon counting is, unsurprisingly, sub-optimal. Despite this, restricting ourselves to photon counting in this regime still offers a quantum advantage compared to the coherent state benchmark. As $\varGamma \rightarrow 1$, the optimal probe tends to a superposition of the lowest-lying even number states, whose QFI across the entire region $\varGamma \gtrsim 0.4$ is somewhat comparable to the maximal ones found through optimisation. Correlations between these non-zero states provide additional contributions to the QFI which render photon counting sub-optimal here. However, in the large TPA limit where they are truly optimal, they become negligible and photon counting is very nearly optimal.

\begin{figure}
    \centering
    \includegraphics[width=0.85\linewidth]{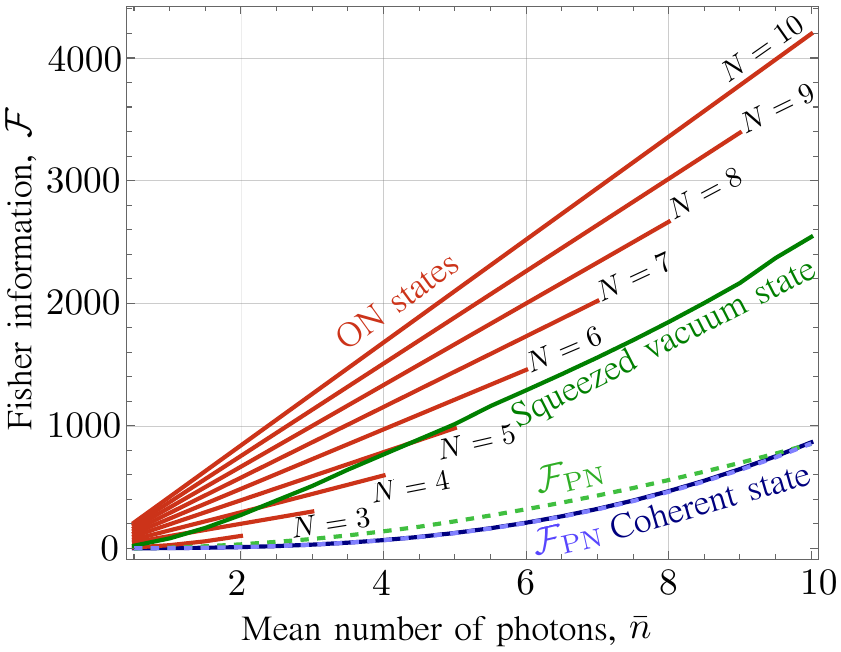}
    \caption{Scaling of QFI (solid), $\mathcal{F}_{{\mathrm{Q}}}$, and FI (dashed) due to photon number measurements, $\mathcal{F}_{\hat{n}}$, with the mean number of photons, $\bar{n}$, for $\varGamma = 0.01$. Results are shown for the coherent and squeezed vacuum states, as well as the ON state $\ket{\mathrm{ON}} = \sqrt{1-\bar{n}/N} \ket{0} + \sqrt{\bar{n}/N} \ket{N}$ which extends from different ON states, labeled with associated $N$-value, to the pure Fock state with $\bar{n} = N$.}
    \label{fig:qfiscaling}
\end{figure}

\subsection{Scaling of Fisher information with mean}

The scaling of both the QFI, $\mathcal{F}_{{\mathrm{Q}}}$, and FI based on photon counting, $\mathcal{F}_{\rm PN}$, with the mean probe energy $\bar{n}$ for each of the quantum states considered is shown in Fig.~\ref{fig:qfiscaling} for $\varGamma = 0.01$. Here, we consider the ON state $\ket{\mathrm{ON}} = \sqrt{1-\bar{n}/N} \ket{0} + \sqrt{\bar{n}/N} \ket{N}$ (labelled by $N$) which extends, with increasing $\bar{n}$, from the ON state to the pure Fock state with $\bar{n}$ for a given $N$. For these states, when $N$ is odd we have $\mathcal{F} = \mathcal{F}_{\mathrm{PN}}$, while when $N$ is even, we have $\mathcal{F} \simeq \mathcal{F}_{\mathrm{PN}}$, with the difference becoming smaller for the larger $N$ values. Known behaviours of coherent and squeezed vacuum states are also recovered. Meanwhile the ON/Fock states show an improvement in TPA parameter estimation over the squeezed vacuum state for some choice of $N$ and there is always a quantum advantage compared to coherent states.

\section{Conclusion}

We have determined that, for small-intermediate absorption, the optimal probe for TPA parameter estimation is the absorption-dependent ON state for a given, fixed mean. That is, a superposition of the vacuum and, for increasing absorbance, progressively decreasing-energy (odd) Fock states, weighted to satisfy the mean constraint. In the limit of large loss, the optimal probe reduces to the pure Fock state when $\bar{n} \in \mathds{E}$, otherwise, a superposition of the lowest-lying even number states. 

Our findings have identified a stark contrast between the QFI behaviours of even and odd Fock states beyond the limit of vanishing absorbance which explains the form of the optimal probe: the latter's QFI dominates in the intermediate regime while in the limit of infinite absorption, tends towards zero. In this region, the QFI of even number states diverge once more, akin to the squeezed vacuum and coherent states, providing further insight into the already-known limiting behaviours of Gaussian probes, namely the coherent and squeezed vacuum state, to which our results are compared. 

Remarkably, when the optimal probe takes the form of an odd ON state or even Fock state, the optimal measurement is simple photon counting.  Otherwise, in the limit of large TPA and for $\bar{n} \notin \mathds{E}$, photon counting forms a very nearly optimal measurement.



Future research could extend this to uncover the effects of one-photon loss, as done in Ref.~\cite{panahiyan2022two} for Gaussian states, arising from potential experimental inefficiencies to determine the practical capabilities of such schemes. Another direction would be the consideration of two-mode problems with ancilla entanglement~\cite{nair2023phasecov} to determine whether additional quantum features can be exploited to further increase sensitivity.

\textbf{Acknowledgments.}~ 
A.K., K.-G.L., and C.L. are supported by Institute for Information \& communications Technology Planning \& Evaluation (IITP) grant funded by the Korea government (MSIT) (RS-2023-00222863). A.K. and C.L. are also supported by the National Research Council of Science \& Technology (NST) grant by the Korea government (MSIT) (No. CAP22000-000). R.N. is supported by the Singapore Ministry of Education Tier 2 Grant No. T2EP50221-0014. A.C. acknowledges the Ministry of Education, Singapore and the National Research Foundation, Singapore and the NRF QEP grant NRF2021-QEP2-02-P03.

\newpage
\appendix

\section{Effect of two-photon absorption on arbitrary quantum states\label{App1}}

Recall from the main text that our TPA model is given by
\begin{equation}
\label{TPAmodelApp}
	\frac{d \hat{\rho}_t}{dt}   = \gamma \, \mathscr{L} \hat{\rho}_t 
	                                \equiv \frac{\gamma}{2} \bigg( \hat{a}^2 \, \hat{\rho}_t \, \hat{a}^\dag{}^2 - \frac{1}{2} \, \hat{\rho}_t \, \hat{a}^\dag{}^2 \hat{a}^2  - \frac{1}{2} \, \hat{a}^\dag{}^2 \hat{a}^2 \, \hat{\rho}_t \bigg)  \; .
\end{equation}
This has been solved exactly using different methods~\cite{simaan1975quantum,simaan1975quantum2,simaan1978off,gilles1993two,klimov2003algebraic}. Here we provide a brief discussion of the solution provided by Ref.~\cite{klimov2003algebraic} in a more transparent notation. When parametrised in terms of $\varepsilon=\gamma\,t$, this gives an output state of the form 
\begin{align}
\label{KlimovSoln}
	\hat{\rho}_\varepsilon = \sum_{l=0}^\infty \sum_{k=l}^\infty \, \frac{(-1)^l}{(k-l)!\,l!} \; \frac{\exp[ - (\varepsilon/4) \, \mathscr{K}_l ]}{\mathscr{J}_{k,l}} \; \hat{a}^{2k} \, \hat{\rho}_0 \, \hat{a}^\dag{}^{2k}  \; , 
\end{align}
for any input state $\hat{\rho}_0$. The superoperators $\mathscr{K}_l$ and $\mathscr{J}_{k,l}$ are defined by
\begin{align}
\label{Ksupop}
	\mathscr{K}_l \, \hat{s} &\equiv  \big\{ (\hat{n}+2l)(\hat{n}+2l-1), \hat{s} \big\}  \; ,  \\
\label{Jsupop}
	\mathscr{J}_{k,l} \, \hat{s} &\equiv \underset{(m \ne l)}{\prod^k_{m=0}} \big\{ 2\hat{n}+2m+2l-1, \hat{s} \big\}  \; ,
\end{align}
where $\{ \hat{r},\hat{s} \} \equiv \hat{r}\,\hat{s} + \hat{s}\,\hat{r}$ for any $\hat{r}$ and $\hat{s}$. Note that $\mathscr{J}_{k,l}$ appears in the denominator of a superoperator fraction in Eq.~\eqref{KlimovSoln}. This denotes a superoperator $\mathscr{J}^{-1}_{k,l}$ such that
\begin{align}
	\mathscr{J}_{k,l}^{-1} \, \mathscr{J}_{k,l} = \mathscr{J}_{k,l}\, \mathscr{J}^{-1}_{k,l} = \mathbbm{1}  \;,   \quad    \big[ \mathscr{J}^{-1}_{k,l}, \mathscr{K}_l \big]=0  \; .
\end{align}
It can be verified from Eqs.~\eqref{Ksupop} and \eqref{Jsupop} that such a $\mathscr{J}^{-1}_{k,l}$ always exists, and the superoperator fraction in Eq.~\eqref{KlimovSoln} is well defined (i.e.~can be read without any ambiguity).

In practice, we find it convenient to rewrite Eq.~\eqref{KlimovSoln} in a slightly different way:
\begin{align}
\label{KlimovSolnRewrite}
	\hat{\rho}_\varepsilon = {}& \sum_{l=0}^\infty \sum_{k'=0}^\infty \, \frac{(-1)^l}{k'!\,l!} \; \frac{\exp[ - (\varepsilon/4) \, \mathscr{K}_l ]}{\mathscr{J}_{l+k',l}} \; \hat{a}^{2(l+k')} \, \hat{\rho}_0 \, \hat{a}^\dag{}^{2(l+k')}  \\
\label{KlimovSolnFinal}
	= {}& \sum_{k=0}^\infty \sum_{l=0}^k \, \frac{(-1)^l}{(k-l)!\,l!} \; \frac{\exp[ - (\varepsilon/4) \, \mathscr{K}_l ]}{\mathscr{J}_{k,l}} \; \hat{a}^{2k} \, \hat{\rho}_0 \, \hat{a}^\dag{}^{2k}  \; .
\end{align}
The first equality simply expresses the sum over $k$ in Eq.~\eqref{KlimovSoln} as a sum over $k'\equiv k-l$. The second equality follows on first reordering the double sum in Eq.~\eqref{KlimovSolnRewrite}, dropping the prime on the dummy index $k'$, and then using the following identity for any summand $S_{k,l}$
\begin{align}
	\sum_{k=0}^\infty \sum_{l=0}^\infty \; S_{k,l} = \sum_{k=0}^\infty \sum_{l=0}^k \; S_{k-l,l}   \;.
\end{align}

The numerical results in the main text have been obtained with Eq.~\eqref{KlimovSolnFinal} where necessary since it provides a straightforward way to reason about the quantum state's dynamics and carry out calculations. This final form limits us to considering only a single infinite sum, which can readily be limited by the input quantum state $\hat{\rho}_0$ in its Fock representation. Also, since the second summation is confined for each $k$, each term in $k$ simply corresponds to the order of the TPA transition, i.e., for $k=1$ it refers to the transition $\ket{n} \rightarrow \ket{n-2}$. It is worth highlighting again that Eq.~\eqref{KlimovSolnFinal} permits us to consider a much larger set of input states for the estimation of TPA rates than had been possible previously.

\section{Photon number distribution and Fisher information}\label{App2}

The FI based on photon counting may be determined directly via the input photon number distributions. In particular, one can make use of an alternative solution to the master equation which involves solving the system of differential equations given by
\begin{align}
    & 4 \, \frac{\partial}{\partial \varepsilon} \bra{n}\hat{\rho}_{\varepsilon}\ket{n'} \nonumber \\
    &= \left[ (n+1)(n+2)(n'+1)(n'+2) \right]^{\frac{1}{2}} \bra{n+2}\hat{\rho}_{\varepsilon}\ket{n'+2} \nonumber \\ &\quad - \frac{1}{2} \left[ n(n-1)+n'(n'-1)\right] \bra{n}\hat{\rho}_{\varepsilon}\ket{n'} .
\end{align}
The FI with PN measurements can now be derived from Eq.~(\ref{eq:CFI}) of the main text. The result is
\begin{equation}
    \mathcal{F}_{\rm PN} = \sum_{n=0}^\infty \frac{\left[(n+1)(n+2)p(n+2|\varepsilon) -n(n-1)p(n|\varepsilon) \right]^2}{4\,p(n|\varepsilon)} 
\end{equation}
where $p(n|\varepsilon)$ is the probability of measuring $n$ photons in the state $\hat{\rho}_\varepsilon$, i.e. 
\begin{align}
    p(n|\varepsilon) = \bra{n}\hat{\rho}_\varepsilon\ket{n} = \Tr\big[\dyad{n}{n}\hat{\rho}_\varepsilon\big] \; .
\end{align}

\section{Scaling of Fock state quantum Fisher information in the limit of low TPA\label{App3}}

To determine the scaling of the QFI/FI for Fock states with photon number $n$, we first note that the QFI takes the simplified form for states diagonal in the Fock basis:
\begin{align}
    \mathcal{F}_{\mathrm{Q}} = {}& \sum_{k=0}^\infty \, \bra{k} \hat{L}^2_\varepsilon \ket{k}\bra{k} \hat{\rho}_{\varepsilon} \ket{k}  \\
    = {}& \sum_{k=0}^\infty \, \frac{1}{h_k(n,\varepsilon)} \, \bigg[ \frac{\partial }{\partial \varepsilon} \, h_k(n,\varepsilon) \bigg]^{\!2}.
    \label{eq:fockQFIapp}
\end{align}
We have defined the function $h_k(n,\varepsilon)$ to be
\begin{align}
    h_k(n,\varepsilon) \equiv &{} \, \sum_{l=0}^{k} \, \frac{(-1)^l}{(k-l)! l!} \, \frac{n!}{(n-2k)!} \; F_{k,l}(n,\varepsilon) \,, 
    \label{eq:hktermsFockapp}
\end{align}
where
\begin{align}
    F_{k,l}(n,\varepsilon) \equiv &{} \,  \frac{\exp\!\big[\!-\varepsilon\,(n-2k + 2l)(n-2k + 2l-1)/2\big]}{ 2\,\underset{(m \ne l)}{\prod^k_{m=0}}(2n-4k+2m+2l-1)} \, .
    \label{eq:hkFockapp}
\end{align}
The only possible contribution to the divergence of the Fock state's QFI, occurring when $h_k(n,\varepsilon)=0$, is from the term arising from a single case of TPA, i.e., $k=1$. The associated term in the summation for an arbitrary Fock state $\hat{\rho}_0 = \dyad{n}{n}$ is given by
\begin{align}
\label{eq:h1term}
    \frac{1}{h_1}\left(\frac{\partial h_1}{\partial \varepsilon} \right)^2 = {}& \frac{ n!}{ 8(2n-3)(n-2)!} \; e^{-(6-n+n^2)\varepsilon/2} \nonumber \\
    & \times \frac{\left[ e^{3\varepsilon} (n-1)n - e^{2n \varepsilon} (6-5n+n^2)\right]^2}{ e^{2n\varepsilon} - e^{3 \varepsilon}}.
\end{align}
Computing a Laurent series for small $\varepsilon$, we then find that
\begin{equation}
  \lim_{\varepsilon \rightarrow 0}\,\frac{1}{h_1}\left(\frac{\partial h_1}{\partial \varepsilon}\right)^2  = \frac{n(n-1)}{2 \varepsilon}.
  \label{eq:finalfockscalingAPP}
\end{equation}

Adapting the same procedure to determine ON state scaling in this same limit, first note that the single case of TPA occurs from the $\ket{N}$ state contribution which now has an extra weight $\bar{n}/N$ associated with it. Then, Eq.~(\ref{eq:finalfockscalingAPP}) becomes
\begin{equation}
  \lim_{\varepsilon \rightarrow 0}\frac{1}{h_1}\left(\frac{\partial h_1}{\partial \varepsilon}\right)^2  = \frac{\bar{n}}{N}\bigg[\frac{N(N-1)}{2 \varepsilon}\bigg] = \frac{\bar{n}(N-1)}{2 \varepsilon}.
  \label{eq:finalONscalingAPP}
\end{equation}

Taking into account the reparameterisation $\varGamma = 1-\exp(-\varepsilon)$ in the limit of $\varGamma \rightarrow 0$ for Eqs.~(\ref{eq:finalfockscalingAPP}) and (\ref{eq:finalONscalingAPP}), it is straightforward to recover Eqs.~(\ref{eq:fockscalingsmallTPA}) and (\ref{eq:ONscalingsmallTPA}) of the main text, respectively.


\section{Comparison between even and odd number state behaviours across TPA scales}\label{AppEvenOdd}

\subsection{Small-intermediate TPA}

Consider as an initial probe the Fock state $\ket{n}$. After TPA evolution, the output state is diagonal in the photon number basis and its QFI takes a simplified form of Eq.~(\ref{eq:fockQFIapp}), outlined in Appendix~\ref{App3}. Noting the form of Eqs.~(\ref{eq:hktermsFockapp}) and (\ref{eq:hkFockapp}), ultimately the QFI is a summation of individual terms over $k \in \big[0,\lfloor N/2 \rfloor \big]$, which indexes the order of the TPA transition, and $l \in [0,k]$. 
 
Each of the $k$ terms in the QFI, corresponding to the constituent number state $\ket{n-2k}$. Among these, the most dominant term in both $k$ and $l$ occurs when $k=l=0$ and is proportional to $\binom{n}{2}^2 (1-\varGamma)^{\binom{n}{2}}$, where we have also taken into account the reparameterisation $\varGamma = 1 - \exp(-\varepsilon)$. Let us now consider the differing behaviours under $n\rightarrow 2n$ and $n\rightarrow 2n+1$, representing even and odd Fock state inputs (both with the same number $k$ of QFI contributions), respectively. Then, this term becomes
\begin{equation}
    \mathrm{even:} \quad \left[n(2n-1)\right]^2 (1-\varGamma)^{n(2n-1)},
\end{equation}
\begin{equation}
    \mathrm{odd:} \quad \left[n(2n+1)\right]^2 (1-\varGamma)^{n(2n+1)}.
\end{equation}
It is clear that odd number state contributions to the QFI initially have a larger value compared to those from even number states for small $\varGamma$, but decay much faster as $\varGamma$ increases. This decay rate is larger still with higher number states, thus as $\varGamma$ increases the value of optimal $N(\varGamma)$ decreases.

\subsection{Large TPA}

Consider now the behaviour seen in the limit of large loss as $\varGamma \rightarrow 1$ for even and odd Fock state QFI contributions. In particular, let us compare the QFI contribution associated with $k=1$, $l=1$ which, in terms of $\varGamma$, is proportional to
\begin{equation}
    \mathrm{even:} \quad (1-\varGamma)^{-2} (1-\varGamma)^{n(2n-1)} \rightarrow (1-\varGamma)^{-1},
\end{equation}
\begin{equation}
    \mathrm{odd:} \quad (1-\varGamma)^{-2} (1-\varGamma)^{n(2n+1)} \rightarrow (1-\varGamma)^{1},
\end{equation}
where the final limiting value is calculated for $n=1$, referring to contributions from $\ket{2}$ and $\ket{3}$ states, respectively. Note that regardless of the initial state, all even and odd number states will, after TPA, end up with populations in these states and their QFI will have a contribution of this form. Considering what happens now with $\varGamma \rightarrow 1$, it is clear that an even number state's QFI will eventually diverge in this limit while an odd number state's QFI will become zero.

\section{Phase insensitivity of the quantum Fisher information}\label{App4}

In this appendix, we justify the claim made in Sec.~\ref{sec:optimalTPA} that the QFI for TPA loss is invariant to the direction of displacement for a coherent state probe, and to the squeezing direction for a squeezed state probe. Denote by $\hat{\rho}_0$ a coherent state with real-valued amplitude or a squeezed vacuum state with real squeezing parameter, as we assumed in Sec.~\ref{sec:optimalTPA}. The choice of a different displacement direction or squeezing axis corresponds to using a probe rotated by some angle $\phi$, given by
\begin{equation}
    \hat{\rho}(\phi) = \mathscr{P}_\phi \, \hat{\rho}_0 \;,
\end{equation}
where for any $\hat{s}$, the rotation (or phase-shift) superoperator is defined by
\begin{align}
\mathscr{P}_\phi \, \hat{s} = e^{-i \phi \hat{n}} \, \hat{s} \, e^{i \phi \hat{n}} \;.
\end{align}
It is straightfoward to see that 
\begin{align}
    \mathscr{P}^{-1}_\phi \hat{s} = e^{i \phi \hat{n}} \, \hat{s} \, e^{-i \phi \hat{n}} \;,
\end{align}
and, for any $\hat{r}$ and $\hat{s}$, 
\begin{align}
\label{RotationProperty1}
    \mathscr{P}_\phi \, \hat{r} \hat{s} = {}& \big(\mathscr{P}_\phi \, \hat{r} \big)\big(\mathscr{P}_\phi \, \hat{s} \big) \;,  \\
\label{RotationProperty2}
    \Tr\!\big[ \hat{s} \, \mathscr{P}_\phi \, \hat{r}  \big] = {}& \Tr\!\big[ \hat{r}\, \mathscr{P}^{-1}_\phi \hat{s} \big]  \; .
\end{align}
Let us denote the output state for an rotated input by
\begin{align}
    \hat{\rho}_\varepsilon(\phi) = e^{\varepsilon \mathscr{L}} \hat{\rho}(\phi) \;.
\end{align}
The output state for an unrotated input is then given by
\begin{align}
    \hat{\rho}_\varepsilon(0) = e^{\varepsilon \mathscr{L}} \hat{\rho}(0) = e^{\varepsilon \mathscr{L}} \hat{\rho}_0 \;.
\end{align}
Note that $\hat{\rho}_\varepsilon(0)$ is exactly what we have referred to as $\hat{\rho}_\varepsilon$ in the main text. The SLD for an output with a rotated input, i.e.~$\hat{\rho}_\varepsilon(\phi)$, is then defined by an $\hat{L}_\varepsilon(\phi)$ such that
\begin{equation}
    \frac{\partial}{\partial \varepsilon} \, \hat{\rho}_{\varepsilon}(\phi)= \frac{1}{2} \, \big[ \hat{\rho}_{\varepsilon}(\phi) \, \hat{L}_{\varepsilon}(\phi) + \hat{L}_{\varepsilon}(\phi) \, \hat{\rho}_{\varepsilon}(\phi) \big].
    \label{eq:appd3}
\end{equation}
Likewise, the SLD for an unrotated input state satisfies,
\begin{equation}
    \frac{d}{d\varepsilon} \, \hat{\rho}_{\varepsilon}(0) = \frac{1}{2} \, \big[ \hat{\rho}_{\varepsilon}(0) \, \hat{L}_{\varepsilon}(0) + \hat{L}_{\varepsilon}(0) \, \hat{\rho}_{\varepsilon}(0) \big],
    \label{eq:appd2}
\end{equation}
where $\hat{L}_\varepsilon(0)$ is precisely $\hat{L}_\varepsilon$ from the main text.

It was shown in Ref.~\cite{chia2020phase} that 
\begin{align}
\label{[P,L]=0}
    \big[ \mathscr{P}_\phi, \mathscr{L}] = 0  \;.
\end{align}
Using Eq.~\eqref{[P,L]=0} we thus have
\begin{align}
    \hat{\rho}_\varepsilon(\phi) = {}& e^{\varepsilon \mathscr{L}} \, \mathscr{P}_\phi \, \hat{\rho}(0)  \\ 
    = {}& \mathscr{P}_\phi \, e^{\varepsilon \mathscr{L}} \hat{\rho}(0) 
    = \mathscr{P}_\phi \, \hat{\rho}_\varepsilon(0)  \;.
\end{align}
Now differentiating respect to $\varepsilon$, then using Eqs.~\eqref{eq:appd2} and \eqref{RotationProperty1} gives
\begin{align}
    \frac{\partial}{\partial \varepsilon} \, \hat{\rho}_\varepsilon(\phi) = {}&\mathscr{P}_\phi \, \frac{d}{d\varepsilon} \, \hat{\rho}_\varepsilon(0) \\
    = {}& \frac{1}{2} \, \big[ \mathscr{P}_\phi \, \hat{\rho}_{\varepsilon}(0) \, \hat{L}_{\varepsilon}(0) + \mathscr{P}_\phi \, \hat{L}_{\varepsilon}(0) \, \hat{\rho}_{\varepsilon}(0) \big] \\
\label{drho/deps}
    = {}& \frac{1}{2} \, \big\{\hat{\rho}_{\varepsilon}(\phi) \, \big[\mathscr{P}_\phi\,\hat{L}_{\varepsilon}(0)\big] + \big[\mathscr{P}_\phi\,\hat{L}_{\varepsilon}(0)\big] \, \hat{\rho}_{\varepsilon}(\phi)] \big\} \; .
\end{align}
Equating Eq.~\eqref{drho/deps} to Eq.~\eqref{eq:appd3} then implies that
\begin{align}
\label{RotatedSLD}
    \hat{L}_\varepsilon(\phi) = \mathscr{P}_\phi \, \hat{L}_\varepsilon(0) \; .
\end{align}
Using Eqs.~\eqref{RotationProperty1}, \eqref{RotationProperty2}, and \eqref{RotatedSLD}, we can then show that $\mathcal{F}_\mathrm{Q}(\phi)\equiv \Tr \big[\hat{\rho}_\varepsilon(\phi)\hat{L}^2_\varepsilon(\phi) \,  \big]$ is in fact independent of $\phi$:
\begin{align}
    \mathcal{F}_\mathrm{Q}(\phi) = {}& \Tr \big[ \,\hat{\rho}_{\varepsilon}(\phi) \{ \mathscr{P}_\phi \, \hat{L}_{\varepsilon}(0) \}^2 \, \big]  \\
    = {}& \Tr \big[ \, \hat{\rho}_{\varepsilon}(\phi) \, \mathscr{P}_\phi \, \hat{L}^2_{\varepsilon}(0) \, \big]  \\
    = {}& \Tr \big[ \, \hat{L}^2_{\varepsilon}(0) \, \mathscr{P}^{-1}_\phi \, \mathscr{P}_\phi  \, \hat{\rho}_{\varepsilon}(0) \, \big]  
    = \mathcal{F}_\mathrm{Q}(0)  \; .
\end{align}
This completes our proof that no loss of generality is incurred by choosing a real-valued amplitude for $\hat{\rho}_0$ if it is a coherent state, or a real-valued squeezing parameter if it is a squeezed state.

\end{document}